\documentclass[prl,twocolumn,nofootinbib]{revtex4}
\usepackage{graphicx}
\usepackage{latexsym}
\def\be{\begin{equation}}
\def\ee{\end{equation}}
\def\bea{\begin{eqnarray}}
\def\eea{\end{eqnarray}}

\begin{document}
\title{Gravitational Leptogenesis and Its Signatures in CMB}
\author{Bo Feng}
\author{Hong Li}
\author{Mingzhe Li}
\author{Xinmin Zhang}
\affiliation{Institute of High Energy Physics, Chinese Academy of
Sciences, P.O. Box 918-4, Beijing 100049, People's Republic of
China}

\begin{abstract}
We study the phenomenologies in astrophysics and particularly in
CMB associated with  the gravitational leptogenesis. Our results
show that future CMB polarization experiments, such as PLANCK and
CMBpol will make a possible test on this class of model for
leptogenesis.
\end{abstract}

\maketitle

\hskip 1.6cm PACS number(s): 98.80.Cq \vskip 0.4cm

The origin of the baryon number asymmetry in the universe remains
a big puzzle in cosmology and particle physics. Sakharov's
original proposal for a dynamical generation of the baryon
asymmetry requires three ingredients \cite{sakharov}: (1) baryon
number violation; (2) $C$ and $CP$ non-conservation; and (3) out
of thermal equilibrium. All of the three conditions above in the
minimal standard model (SM) of the particle physics can be
realized, however, quantitatively the baryon asymmetry generated
during the electroweak phase transition is too small to account
for the value observed \cite{manton}: $n_B/s\sim 10^{-10}$. In the
SM, the $CP$ violation is not enough and the first order phase
transition is too weak. Thus far in the literature many models
involving new physics, such as multi-Higgs models, left-right
symmetric models, supersymmetric models and models with extra
dimension have been proposed for baryogenesis.

Theoretically there exist many possible ways to go beyond the SM.
Experimentally a strong evidence for new physics beyond the SM
comes from the establishment on the atmospheric and the solar
neutrino oscillation with an additional support from the reactor
antineutrino, which demonstrates that the neutrinos have masses
and mix with each other. Within the content of the particles in
the SM the neutrino masses and mixing can be described by a
dimension-5 operator
\begin{eqnarray}\label{lepvio}
{\cal L}_{\not L} = \frac{2} { f } l_L l_L \phi \phi +{\rm H.c.}~,
\end{eqnarray}
where $f$ is the scale of new physics beyond the Standard Model,
 $l_L$ and $\phi$ are the
left-handed lepton and the Higgs doublets respectively. When the
Higgs field gets a vacuum expectation value $< \phi > \sim v $,
the left-handed neutrino receives a Majorana mass $m_\nu \sim
\frac{v^2}{f}$.

With the operator in (\ref{lepvio}) the baryon minus lepton
($B-L$) number is violated. It would be very economical if the
leptogenesis \cite{leptogenesis} happens in this minimal extension
of the SM. However the Sakharov's third condition can not be
realized. In the traditional version of the leptogenesis the heavy
right-handed neutrinos are introduced and their non-thermal
equilibrium decays, coupled with the electroweak sphaleron
process, generate the required baryon number asymmetry. In general
at least two types of the right-handed neutrinos are needed for a
successful leptogenesis.

Note that the Sakharov's third condition applies for models where
the $CPT$ is conserved. If the $CPT$ is violated the baryogenesis
or leptogenesis could happen in thermal equilibrium \footnote{For
recent relevant works see, e.g., \cite{li,trodden}.}, such as the
spontaneous baryogenesis \cite{cohenkaplan} and the gravitational
baryogenesis\cite{steinhardt}. In the original version of the
spontaneous baryogenesis by Cohen and Kaplan \cite{cohenkaplan} it
requires an extra scalar field beyond the SM and its derivative
coupling to the baryon current. In models of gravitational
baryogenesis\cite{steinhardt} \cite{lihong} \cite{shiromizu} the
scalar field in \cite{cohenkaplan} is replaced by a function of
the Ricci scalar,
  \be\label{genint}
 {\cal L}_{int} = c ~\partial_{\mu}f(R) J^{\mu}~,
 \ee
where $J^\mu$ is a vector current made of the particles in the
standard model, $f(R)$ a dimensionless function of the Ricci
scalar $R$ and $c$ is the coupling constant characterizing the
strength of this interaction. In Ref.\cite{steinhardt}
$f(R)=R/M^2$ is taken, however the Einstein equation, $R=8\pi G
T^{\mu}_{\mu}=8\pi G (1-3w)\rho$, tells us that $\dot f(R)=0$ in
the radiation dominated epoch.
In
Ref.\cite{lihong} we have proposed a model of gravitational
leptogenesis
\be
 \frac{n_{B-L}}{s}\sim c \frac{\dot f(R) }{T}~,
 \ee
where $f(R)\sim \ln R$, so the term
$\partial_{\mu}f(R)\sim
\partial_{\mu}R/R$ does not vanish during the radiation dominated
epoch and the observed baryon number asymmetry can be generated
naturally via leptogenesis.
Since
 in this scenario of baryogenesis, no extra degrees of freedom beyond
 the SM and classical gravity are introduced we
call it the minimal model of baryogenesis.

The key ingredient in this type of leptogenesis is
the cosmological $CPT$ violation caused by the non-vanishing
${\dot f}(R)$. It would be very interesting to ask if this type of
$CPT$ violation could be tested in the experiments. Unfortunately
in the laboratory the predicted $CPT$ violation as shown in
\cite{lihong} is much below the current experimental sensitivity.
In this paper we will show that the phenomenon associated with the
$CPT$ violation can be tested in the future cosmic microwave
background (CMB) polarization experiments, such as
PLANCK\cite{Planck} and CMBpol\cite{CMBpol}. For this purpose, we
take specifically the $J^\mu$ in (2) to be the Chern-Simons term
of the electromagnetic field, \be\label{couple}
\mathcal{L}_{int}=-\frac{1}{2}\delta~\partial_{\mu}f(R)K^{\mu}~,
\ee where $K^{\mu}=A_{\nu}\widetilde{F}^{\mu\nu}$, $F_{\mu\nu}$ is
the electromagnetic field strength tensor, and
 $\widetilde{F}^{\mu\nu}\equiv {1\over
2}\epsilon^{\mu\nu\rho\sigma}F_{\rho\sigma}$ is its dual. In
(\ref{couple}) $\delta$ is a constant characterizing the strength
of this type of interaction and will be calculable if the
underlying fundamental theory is known. In the framework of
effective theory, we expect it by a naive dimensional analysis to
be suppressed by a factor of $\frac{e^2}{4 \pi^2} $ in comparison
with the constant $c$ in (3). The factor $- \frac{1}{2}$ in (4) is
introduced for the convenience of the discussions.

The term in (\ref{couple}) is gauge invariant and parity-odd. And
it can lead to the rotation of electromagnetic waves when
 propagating over cosmological distances \cite{jackiw}. This effect is
 known as ``cosmological birefringence". By observing the change $\Delta
\chi$
 in position angle $\chi$ of the plane of polarized radiation from distant
 radio galaxies and quasars at a redshift $z$ due to the
 phase shift between the modes of opposite helicity, cosmological
 birefringence is directly observable and has been used to
 constrain the amplitude of Lorentz and parity-violating
 modifications to electrodynamics \cite{jackiw,carroll}.
Refs.\cite{kamionkowski,lepora,Pogosian,brandenberger,Maity,knox95,zspola,kpola}
have considered some effects of the cosmological birefringence in
CMB.

We start our discussion with an examination on the
 experimental limits from quasars on the coupling constant $\delta$ in
 Eq. (\ref{couple}). We will show that, for $f(R)\sim \ln R$, the
 experimental constraints on $\delta$ are not much stringent,
e.g., $|\delta|\lesssim
 {\cal O}(10^{-1})$. Then we will study the effects of our
 model on the polarization of CMB. Our results show that for
  $|\delta| \gtrsim {\cal O}(10^{-7})$, the interaction in Eq.
 (\ref{couple}) will give rise to effects observable in the
future CMB
 polarization experiments.

With the interaction in Eq. (\ref{couple}), the Lagrangian for the
electromagnetic field in the absence of source is \be
 \mathcal{L}_{em}=-{1\over
 4}F_{\mu\nu}F^{\mu\nu}+\mathcal{L}_{int}~.
 \ee
The equations of motion for the electromagnetic field are \be
\nabla_{\mu}F^{\mu\nu}=\delta~\partial_{\mu}f\widetilde{F}^{\mu\nu}~,
\ee and \be \nabla_{\mu}\widetilde{F}^{\mu\nu}=0~, \ee
 where $\nabla_{\mu}$ denotes the covariant
derivative with the metric $g_{\mu\nu}$. In the spatially flat FRW
cosmology, the metric can be written as \be
 ds^2=a^2(\eta)(d\eta^2-\delta_{ij}dx^idx^j)~,
 \ee
where $\eta$ is the conformal time which is related to the cosmic
time by $d\eta=dt/a$. With the conventions,
$A^{\mu}=(A^0,~\bf{A})$ and \bea\label{eb}
 & &\bf{E}=-\nabla A^0-\frac{\partial \bf{A}}{\partial
 \eta}~,\nonumber\\
 & &\bf{B}=\nabla\times \bf{A}~,
 \eea
we can write the electromagnetic field strength tensor in terms of
$\bf{E}$ and $\bf{B}$:
 \be F^{\mu\nu} =a^{-2}\left[\begin{array} {cccc} 0 &-E_x & -E_y & -E_z \\ E_x & 0 & -B_z & B_y \\
 E_y & B_z & 0 & -B_x \\ E_z &-B_y & B_x & 0 \\ \end{array}\right]~. \ee
In Eq. (\ref{eb}), $\nabla$ and $\nabla\times$ represent the usual
gradient and curl operators in the Cartesian three dimensional
space. The dual tensor $\widetilde{F}^{\mu\nu}$ can be obtained
from $F^{\mu\nu}$ by replacing $\bf{E}$ and $\bf{B}$ with $\bf{B}$
and $-\bf{E}$ respectively. In terms of the notations given by
Carroll and Field \cite{field}: \bea && \bf{B}(\vec{x},\eta)={\rm
e}^{-i\bf{k}\cdot\bf{x}}\bf{B}(\eta)~,\nonumber\\
& &F_{\pm}=a^2B_{\pm}(\eta)=a^2(B_y\pm iB_z)~, \eea  we have the
equation of motion for a given mode $\bf{k}$, \be
 F_{\pm}''+(k^2\pm \delta k f')F_{\pm}=0~,
 \ee
 where the prime represents the derivative with respect to $\eta$ and
 $k$ is the modulus of $\bf{k}$.
 In the equation above, we have assumed
the wave vector $\bf{k}$ is along the $x$ axis, and $+$ and $-$
denote the right- and left-handed circular polarization modes
respectively.
  The non-vanishing
 $f'(R)$ induces some difference between the dispersion relations for
 the modes with different handedness.
 This will rotate the direction of the polarization of light from
 distant sources. For a source at a redshift $z$, the
 rotation angle is
 \be
 \Delta \chi=\frac{1}{2}\delta~\Delta f(R)~,
 \ee
where $\Delta f$ is the change in $f$ between the redshift $z$ and
today, i.e., $\Delta f=\left. f \right|_{z}-\left. f
 \right|_{z=0}$.

For $f(R)\sim \ln R$ and in the spatially flat $\Lambda$CDM model,
we get  \be
 \Delta f=\ln (\frac{\Omega_{m0}(1+z)^3+4\Omega_{\Lambda
 0}}{\Omega_{m0}+4\Omega_{\Lambda 0}})~.
 \ee
The subscript $0$ denotes today's value. The underlying model
parameters we set below are from Ref. \cite{Tegmark}: \be (\tau,
\Omega_{\Lambda}, \Omega_d h^2, \Omega_b h^2, A_S, n_S) = (0.17,
0.72, 0.12, 0.024, 0.89, 1) ~.
 \ee
Despite the fact argued in the literature \cite{cb,leahy,cohen}
whether cosmic birefringence has been detected through distant
radio galaxies and quasars, using the data given by Leahy
\cite{leahy} we can get a conservative(see also \cite{cohen})
limit on the coefficient $\delta$ in our model. As an illustrative
effect, in Fig.1 we show three resulting $\Delta \chi- z$ effects
for different values of $\delta$. We can see from Fig.1 that the
single source 3C9 at $z=2.012$, which reads $\Delta \chi=
2^{\circ}\pm 3^{\circ}$ and is consistent with the detailed
analysis of \cite{cohen}, restricts stringently the amplitude of
$\delta$ to be no more than the order of $0.1$. Furthermore, in
order to have an estimation on $|\delta|$, we fit our model to the
data with $\Omega_m= 0.3\pm 0.08$, which is around the 2$\sigma$
limit of the 6-parameter global fit. The corresponding $2 -$
dimensional and $1 -$ dimensional plots with 1 and 2$\sigma$
$C.L.$ are delineated in Fig.2. We get at 1$\sigma$ limit
$\delta=0.03\pm 0.07$ from the $1 -$ dimensional case.

\begin{figure}[htbp]
\begin{center}
\includegraphics[scale=0.6]{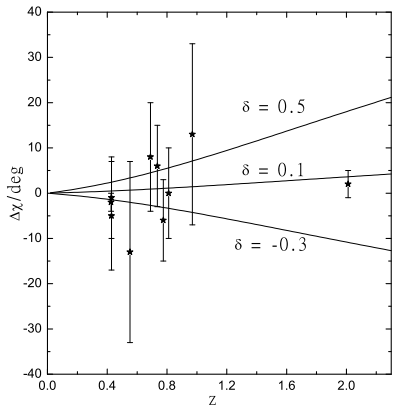}
\caption{Different effects for choosing three sets of $\delta$ of
our model in light of the data from distant radio sources from
\cite{leahy}. \label{fig:fig1}}
\end{center}
\end{figure}

\begin{figure}[htbp]
\begin{center}
\includegraphics[scale=0.6]{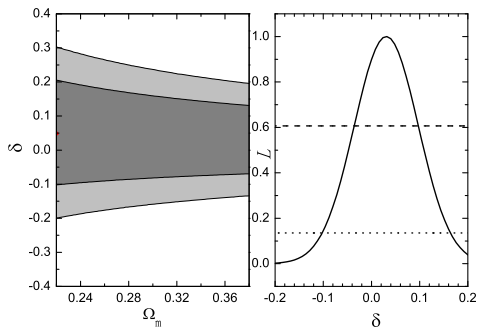}
\caption{ $2 -$ dimensional and $1 -$ dimensional constraints on
$\delta$ in light of the data from distant radio sources from
\cite{leahy}, with 1 and 2$\sigma$ $C.L.$ shown respectively.
\label{fig:fig2}}
\end{center}
\end{figure}

Now we consider the effects of the interaction (\ref{couple}) in
CMB. As is well known the CMB has become the most remarkable tool
for probing the distant past and present status of our universe.
In particular, DASI \cite{DASI} has marked the first detection of
G-mode polarization and TG correlation of CMB. The first year WMAP
measurement \cite{Bennett} has been up to date the most powerful
observation of cosmology, both in deriving very tight constraints
on cosmological parameters and revealing new signatures of our
universe. The polarization of CMB helps particularly to constrain
the reionization depth \cite{Kogut,wmap} and verify the existence
of early cosmological inflation \cite{Peiris} in high precision
surveys like WMAP. Cosmological birefringence rotates the gradient
type polarization fields into the curl type during the
propagations of the CMB photons from the last scattering surface
to the observer, leading to distinctive effects in the CMB
polarization surveys \cite{HHZ97,SF97,kamionkowski} which leaves
hope for being directly detected in high precision CMB
measurements like PLANCK \cite{Planck} and CMBpol \cite{CMBpol}.

The CMB polarization can be described by two Stokes parameters: Q
and U, which can be spherically expanded to get a gradient (G) and
a curl (C) component \cite{HW97}. If the temperature/polarization
distribution does not violate parity, one gets vanished CMB TC and
GC due to the intrinsic properties of the tensor spherical
harmonics. In the case of $P$ violation effect which leads to
cosmological birefringence, the polarization vector of each photon
is rotated by an angle $\Delta \chi$ everywhere and one would get
nonzero TC and GC correlations with \be C_l^{TC}= C_l^{TG}\sin 2
\Delta \chi~, C_l^{GC}= \frac{1}{2}(C_l^{GG}- C_l^{CC})\sin 4
\Delta \chi ~,
 \ee
even though they vanish at the surface of the last scattering.

To model the CMB polarization experiments we follow the notations
of Refs. \cite{knox95,zspola,kpola}. For a full sky pixelized map
with a Gaussian beam with full width at half maximum
$\theta_{FWHM}$, we assume uncorrelated errors of pixels with
uniform variance $\sigma_T^2$ and $\sigma_P^2$ respectively for
temperature and polarization measurements. Assuming negligible
gravitational wave contributions, we show in Fig. 3 the smallest
$\delta$ detectable at $1\sigma$ level using CMB TC signature only
in the left panel and using CMB GC signature in the right panel
with given $\theta_{FWHM}$ and $\sigma_T$. Fig. 3 shows CMB GC
polarization, which has not been studied quantitatively before,
serves more efficient than TC measurements due to different
intrinsic properties of the curl and gradient polarizations.
PLANCK can detect a CMB GC signature for $|\delta|\sim 10^{-5}$
while the $|\delta|$ as small as $\sim 10^{-7}$ is detectable in
CMBpol experiment. We point out here that our result applies to a
general class of cosmological birefringence and  CMB GC signature
in CMBpol can detect $\Delta \chi
> 0.0001 ^{\circ}$.

\begin{figure*}[htbp]
\begin{center}
\includegraphics[scale=0.6]{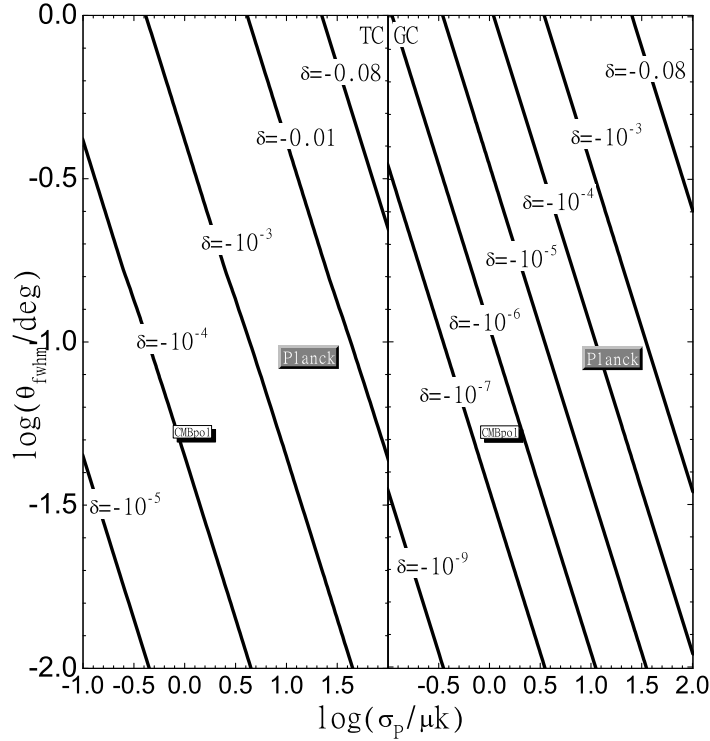}
\caption{The smallest $\delta$ detectable at $1\sigma$ level using
CMB TC and GC signature with given experimental parameters. The
filled and hatched areas illustrate the mission characteristics of
PLANCK and CMBpol respectively. \label{fig:fig3}}
\end{center}
\end{figure*}

Before concluding we point out that this study about the
sensitivity on the coefficient in the interaction (\ref{couple})
from the future CMB measurements has an interesting implication on
the minimal model of the baryogenesis \cite{lihong}. Given Eq. (3)
and $f(R) \sim \ln R$, the final $B-L$ number asymmetry generated
is \bea\label{result} \left.\frac{n_{(B-L)}}{s}\right|_{T_D}\simeq
0.1c\frac{T_D}{m_{pl}}~, \eea where $T_D$ is the decoupling
temperature determined by the $B-L$ violating interaction in
(\ref{lepvio}).

In the early universe the $B-L$ violating rate induced by the
interaction in (\ref{lepvio}) is \cite{sarkar}
\begin{eqnarray}
  \Gamma_{\not L} \sim
    0.04~ \frac{T^3}{ f^2 }~.
\end{eqnarray}
Since $\Gamma_{\not L}$ is proportional to $T^3$, for a given $f$,
$B-L$ violation will be more efficient at high temperature than at
low temperature. Requiring this rate to be larger than the
universe expansion rate $H\sim 1.66 g_{\ast}^{1/2}T^2/ m_{pl}$
until the temperature $T_D$, we obtain a $T_D$-dependent lower
limit on the neutrino mass:
\begin{eqnarray}\label{neutrinomass}
   \sum_i m_i^2  = ( 0.2 ~{\rm eV} ( { \frac{10^{12}~{\rm GeV}}{T_D}
})^{1/2})^2~.
\end{eqnarray}

The experimental bounds on the neutrino masses come from the
neutrino oscillation experiments and the cosmological tests. The
atmospheric and solar neutrino oscillation experiments
give\cite{atmospheric,solar}: \bea & & \Delta
m_{atm}^2=(2.6\pm 0.4)\times 10^{-3} {\rm eV}^2~,\\
& & \Delta m_{sol}^2 \simeq (7.1^{+1.2}_{-0.6})\times 10^{-5} {\rm
eV}^2~. \eea The cosmological tests provide the limits on the sum
of the three neutrino masses, $\Sigma\equiv \sum_i m_i$. The
analysis of WMAP \cite{wmap} and SDSS \cite{Tegmark} show the
constraints: $\Sigma<0.69$ eV and $\Sigma<1.7$ eV respectively.

For the case of normal hierarchy neutrino masses, $m_3\gg m_2,~m_1
$, one has \be
 m_3^2-m_2^2=\Delta m_{atm}^2~,~~~ m_2^2-m_1^2=\Delta m_{sol}^2~,
\ee and \be \sum_i m_i^2\simeq m_3^2\gtrsim \Delta m_{atm}^2. \ee
We can see from Eq. (\ref{neutrinomass}) that this requires the
decoupling temperature $T_D\lesssim 1.5\times 10^{13}$ GeV. For
neutrino masses with inverted hierarchy, $m_3\sim m_2\gg m_1$, we
get \be
 m_3^2-m_2^2=\Delta m_{sol}^2~,~~~ m_2^2-m_1^2=\Delta m_{atm}^2~,
\ee and
 \be \sum_i m_i^2\simeq 2m_3^2\gtrsim 2\Delta m_{atm}^2. \ee
It constrains the decoupling temperature as $T_D\lesssim
7.7\times10^{12}$ GeV. If three neutrino masses are approximately
degenerated, ~{\rm i.e.}~, $m_1 \sim m_2 \sim m_3\sim {\bar m}$,
one has $\Sigma = 3 {\bar m}$ and $\sum_i m_i^2 \simeq
\Sigma^2/3$. In this case, the WMAP and SDSS data require $T_D$ to
be larger than $2.5\times 10^{11}$ GeV and $4.2\times 10^{10}$ GeV
respectively. So, for a rather conservative estimate, we consider
$T_D$ in the range of $10^{10} {\rm GeV}\lesssim T_D \lesssim
10^{13} {\rm GeV}$. Combined with Eq. (\ref{result}) for a
successful leptogenesis this results in a constraint on the
coupling constant $c$: $|c|\gtrsim 10^{-3}$. As argued above the
dimensional analysis indicates $\delta\sim\frac{e^2}{4\pi^2} c$.
The upper limit on $|c|$ for a successful baryogenesis implies
that $|\delta|\gtrsim 10^{-6}$ which as we show above lies in the
range sensitive to the future CMB measurements.

In summary, we have in this paper studied the phenomenon
associated with the gravitational baryo(lepto)genesis in CMB. In
this type of models for baryo(lepto)genesis no extra degrees of
freedom beyond the standard model of particle physics and
classical gravity are introduced, however the effective operators
involving the derivative couplings between the standard model
particles and the gravity play an important role in the generation
of baryon number asymmetry. These operators give rise to CPT
violation during the evolution of the Universe. In this paper we
have considered explicitly the effects of the effective operator
(4) in CMB and showed that the future CMBpol experiments can test
this type of effective interaction for $|\delta|$ as small as
$10^{-7}$. Our results have an interesting implication in
gravitational baryogenesis and show a possible way to test this
type of baryogenesis in the
 future CMB polarization experiments, such as
PLANCK and CMBpol.

{\bf{Acknowledgments:}} We used a modified version of
CMBfastf\cite{cmbfast,IEcmbfast}. We thank Marc Kamionkowski for
enlightening discussions on CMB polarizations and comments on the
manuscript. We thank Robert Brandenberger and Hiranya Peiris for
comments on the paper. We are also grateful to  Xiao-Jun Bi and
Peihong Gu for useful conversations. This work is supported in
part by National Natural Science Foundation of China under Grant
Nos. 90303004 and 19925523 and by Ministry of Science and
Technology of China under Grant No. NKBRSF G19990754.

{}

\end{document}